\documentstyle[12pt,aaspp4,flushrt]{article}

\def\etal{{\it et~al.~}}

\begin{document}

\title{The Centers of Early-Type Galaxies with HST II: Empirical
Models and Structural Parameters\footnote[1]{Based 
on observations with the NASA/ESA Hubble Space Telescope, obtained at the
Space Telescope Science Institute, which is operated by AURA, Inc., under
NASA contract NAS 5-26555.}}

\author{Yong-Ik Byun\footnote[2]{Present address: Institute of
Astronomy, National Central University, Chung-Li, Taiwan 32054,
R.O.C.}}
\affil{Institute for Astronomy, University of Hawaii, 2680
Woodlawn Dr., Honolulu, HI 96822}

\author{Carl J. Grillmair\footnote[3]{Present address: Jet Propulsion Laboratory, MS 183-900, 4800 Oak Grove Dr., Pasadena, CA 91109}}
\affil{UCO/Lick Observatory, University of California, Santa Cruz,
CA 95064}

\author{S. M. Faber}
\affil{UCO/Lick Observatory, Board
of Studies in Astronomy and Astrophysics,
University of California, Santa Cruz, Santa Cruz, CA 95064}

\author{Edward A. Ajhar}
\affil{Kitt Peak National Observatory,
National Optical Astronomy Observatories,\footnote[4]{Operated by AURA under cooperative agreement with the
National Science Foundation.}\\
P. O. Box 26732, Tucson, AZ 85726}

\author{Alan Dressler}
\affil{The Observatories of the Carnegie Institution of Washington\\
813 Santa Barbara St., Pasadena, CA 91101}

\author{John Kormendy}
\affil{Institute for Astronomy, University of Hawaii,
2680 Woodlawn Dr., Honolulu, HI 96822}

\author{Tod R. Lauer}
\affil{Kitt Peak National Observatory,
National Optical Astronomy Observatories,\footnotemark[4]\\
P. O. Box 26732, Tucson, AZ 85726}

\author{Douglas Richstone}
\affil{Department of Astronomy, University of Michigan,
Ann Arbor, MI 48109}

\author{Scott Tremaine}
\affil{Canadian Institute for Theoretical Astrophysics,
University of Toronto\\ 
60 St. George St., Toronto, M5S 1A7, Canada}

\begin{abstract}

We present a set of structural parameters for the central parts of 57
early-type galaxies observed with the Planetary Camera of the Hubble
Space Telescope.  These parameters are based on a new empirical law
that successfully characterizes the centers of early type galaxies.
This empirical law assumes that the surface brightness profile is a
combination of two power laws with different slopes $\gamma$ and
$\beta$ for the inner and outer regions.  Conventional structural
parameters such as core radius and central surface brightness are
replaced by break radius $r_b$, where the transition between power-law
slopes takes place, and surface brightness $\mu_b$ at that radius. An
additional parameter $\alpha$ describes the sharpness of the break.
The structural parameters are derived using a $\chi^2$ minimization
process applied to the mean surface brightness profiles.  The
resulting model profiles generally give very good agreement to the
observed profiles out to the radius of $\sim 10^{\prime\prime}$ imaged
by the Planetary Camera.  Exceptions include galaxies which depart
from pure power-laws at large radius, those with strong nuclear
components, and galaxies partly obscured by dust.  The uncertainties
in the derived parameters are estimated using Monte-Carlo simulations
which test the stability of solutions in the face of photon noise and
the effects of the deconvolution process.  The covariance of the
structural parameters is examined by computing contours of constant
$\chi^2$ in multi-dimensional parameter space.

\end{abstract}

\keywords{galaxies: nuclei - galaxies: elliptical and lenticular, cD}

\section{Introduction}

	Recent Hubble Space Telescope (HST) observations of early-type
galaxies reveal that the central parts have surface-brightness
distributions that are different from the extrapolation of traditional
fitting formulae derived from ground-based observations (Crane {\it et
al.} 1993; Kormendy {\it et al.} 1995; van den Bosch {\it et al.}
1994; Forbes, Franx, \& Illingworth 1995, Lauer {\it et al.} 1995,
hereafter Paper I).  The surface brightness profiles generally consist
of two distinct regions; a steep power-law regime ($I(r) \propto
r^{-\beta}$) at large radius, and a shallower power law ($I(r)\propto
r^{-\gamma}$) at small radius.  Galaxies with a quite shallow inner
power-law slope ($\gamma <$ 0.3) are classified as ``core'' galaxies,
while those with $\gamma > 0.5$ are labeled power-law galaxies (Paper
I). Note that $\gamma$ is significantly different from zero in most
objects studied with HST thus far, with only rare exceptions
(van den Bosch {\it et al.} 1994; Paper I; Ajhar {\it et al.} 1995).

Conventional parameters characterizing the luminosity structure of the
central regions are the central surface brightness and the core
radius, defined as the radius where the surface brightness is half the
central value.  These parameters have been very useful in
understanding elliptical galaxies and bulges of later type galaxies,
especially through the fundamental plane correlations (Kormendy 1982,
1984, 1985, 1987a, b; Lauer 1985).  However, as we
do not find constant surface brightness cores in most HST 
observations, we set about finding an alternative
parameterization (presented below) to better describe the observed
surface brightness distributions.  For a more detailed history on this
subject, readers are referred to Paper I.

	An alternative approach to the one taken here is that of
Gebhardt {\it et al.} 1996), who carry out a non-parametric analysis
of essentially the same sample considered here. The Gebhardt {\it et
al.} analysis is both more thorough and more accurate, but it largely
verifies the simple fits we describe below. The major advantage
of the approach taken in the present paper is that it reduces each
galaxy to a set of 5 numbers. On the other hand, the non-parametric
approach does not force galaxies into possibly ill-fitting boxes.

A one-parameter family of models for spherical stellar systems which
yields double power-law surface brightness profiles was introduced
independently by Dehnen (1993; $\gamma$ models) and by Tremaine \etal
(1994; $\eta$ models). In the notation of Tremaine \etal, the density

\begin{equation}
\rho_\eta(r) \equiv {\eta \over 4 \pi}{1 \over
r^{3-\eta}(1+r)^{1+\eta}}, \hspace{1cm}0 < \eta \le 3.
\end{equation}

However, the majority of the actual galaxy profiles is not well-fit
by $\eta$-models.  Although these models allow for a range of inner
power-law slopes, the variations apparent in the sharpness of the
transition from inner to outer power-laws cannot be reproduced, and
the fixed asymptotic outer slope of $r^{-3}$ does not fit all
galaxies.

To parametrize the HST-observed surface brightness profiles, we introduced in
Paper I a more general
empirical double power-law (the ``Nuker'' law):

\begin{equation}
 I(r) = I_b ~2^{\frac{\beta-\gamma}{\alpha}}
	  \left( \frac{r}{r_b}\right) ^{-\gamma} 
	  \left[ 1 +  \left( \frac{r}{r_b} \right) ^ \alpha \right] ^
	  {- \frac{\beta-\gamma}{\alpha}} 
\end{equation}

\noindent{}The break radius, $r_b$, is the radius at which the steep
outer profile, $I(r)\propto r^{-\beta}$, ``breaks'' to become the
inner shallow profile, $I(r) \propto r^{-\gamma}$.  More precisely,
$r_b$ is the radius at which the absolute logarithmic curvature in the surface
brightness profile, $|d^2\log I(r) /d(\log r)^2|$, is a maximum.
$I_b$ is the vertical scaling parameter and is simply the surface
brightness at $r_b$. As shown in Paper I, the parameter $\alpha$ is
necessary to account for the variations in the
sharpness of the break.

The Nuker law contains many simpler fitting formulae as special cases;

\begin{itemize}

\item The Hubble-Reynolds law (Reynolds 1913; Hubble 1930),	
$$I(r)={I_0a^2\over (r+a)^2},$$
corresponds to $\alpha=1$, $\beta=2$, $\gamma=0$.

\item The modified Hubble or analytic King law (Binney \& Tremaine 1987),
$$I(r)={I_0r_c^2\over r_c^2+r^2},$$
corresponds to $\alpha=2$, $\beta=2$, $\gamma=0$.

\item de Vaucouleurs' $r^{1/4}$ law (de Vaucouleurs 1948),
$$I(r)=I_0\exp(-kr^{1/m}), \qquad m=4,$$
and its generalization to arbitrary $m$ (Sersic 1968),
corresponds to the parameters $\alpha={1\over m}$, $\gamma=0$, in the limit
$r_b\to\infty$, $\beta={1\over m}kr_b^{1/m}\to\infty$.

\item  Ferrarese et al. (1994) fit HST photometry of early-type 
galaxies to the law,
$$I(r)=2^{1/2}I_c\left(r\over r_c\right)^{-\beta_1}\left[1+\left(r\over
r_c\right)^{2(\beta_2-\beta_1)}\right]^{-1/2},$$
which corresponds to $\alpha=2(\beta_2-\beta_1)$, $\beta=\beta_2$,
$\gamma=\beta_1$.

\end{itemize}

Galaxies in which the three-dimensional luminosity density varies
smoothly near the center are said to have ``analytic cores''.  An
analytic core requires that the surface brightness near the center has
the form $I(r)=I_0+ I_1 r^2 + \hbox{O} (r^4)$.  Thus only galaxies
with $\alpha=2$, $\gamma=0$ have analytic cores.

The Nuker law also provides excellent fits to the family of $\eta$-models
(Figure 1).  The behavior of the five Nuker parameters for each of the
$\eta$-models is illustrated in Figure 2 in order to show the parameter
space occupied by $\eta$-models.  As will be seen below, real galaxies cover
a much wider range in these parameters.

The Nuker law is convenient not only for describing the observed surface
brightness distribution but also for computing quantities such as the
luminosity density and the local and line-of-sight velocity dispersion
distributions.  A FORTRAN code that carries out these tasks is available on
request from S. Tremaine.

In this paper, we fit the Nuker law and derive the associated
structural parameters for 57 early-type galaxies observed with the
HST. With the exception of NGC 4881, all objects were observed prior
to the HST refurbishment with the Planetary Camera ($0.\prime\prime043$
per pixel) using filter F555W (corresponding roughly to Johnson $V$).
The sample includes early-type galaxies observed as part of our Guest
Observer (GO) programs (\#2600, \#3912, PI Faber), as part of several
WFPC Instrument Definition Team programs (\#1105, \#3229, \#3286,
\#3639, \& \#5233, PI Westphal), and includes F555W PC images of
early-type galaxies publically available in the HST archive by June
23, 1994. The latter were taken as part of GO programs \#1038 (PI
Ford) and \#2607 (PI Jaffe).  While small amounts of dust could
generally be avoided in determining surface brightness profiles,
galaxies with severe nuclear dust obscuration were rejected. Two or
more exposures were used for each galaxy to permit cosmic ray
rejection, and the peak counts per pixel in the coadded frames were
typically $\sim$2000 before deconvolution. Further details of data
acquisition and reduction of the GO \#2600/3912 sample are given in
Paper I. These reduction procedures have been uniformly applied to the
additional galaxies in the present sample. The only exception is NGC
4881, which was observed after the refurbishment of HST and therefore
did not require extensive deconvolution.  The parameters
determined in this paper are used in the analysis and interpretation
of core properties presented in Faber {\it et al.}  (1996, Paper III).

\section{Fitting Scheme}

The five structural parameters are determined using a
$\chi^2$-minimization method.  $\chi^2$ is defined as the sum of the
squares of the differences in surface brightness, $\Delta \mu(r)^2$,
between the observed profiles and those generated by the Nuker law,
where $\mu(r)$ is in units of magnitudes per arcsecond squared. We
make use of the {\it mean} profiles rather than the major-axis
profiles published in Paper I.  Each galaxy is fitted with a serious
of concentric ellipses and the {\it mean} profile is defined so that
for each isophotal ellipse, $r = \sqrt{r_{maj} \cdot r_{min}}$, where
$r_{maj}$ and $r_{min}$ are major and minor axis radii, respectively.
The use of mean profiles rather than major-axis profiles reduces the
complexities caused by the variation of ellipticity with radius.
Significant variations in ellipticity may cause the apparent shape of
the major axis and mean surface brightness distributions to differ
quite substantially.

As discussed in Paper I, the surface brightness measurements were made
using two procedures, one optimized for high resolution in the inner
arcsecond (Lauer 1985) and the other optimized to handle the low
signal-to-noise ratio data at larger $r$. The two profiles were
subsequently combined.  However, the resulting profile has
inappropriate spacing between data points when plotted in logarithmic
space; this is especially true for the profile obtained from the high
resolution optimized technique, for which the point spacing is linear.
In order to treat all radii with equal importance for our power-law
fitting, the observed profiles were resampled beginning at the first
measured data point ($0.044\arcsec/\sqrt{2\pi} = 0.017\arcsec$) and
progressing at equal logarithmic intervals of $\Delta \log r=0.05 $
using spline interpolation.  This spacing is small enough to represent
the variation of surface brightness in the original profiles.  Uniform
weights and errors were applied to the individual data points in the
resampled profile during the fitting procedure.

For the minimization we have used the MINUIT package, which handles
non-linear minimization through a combination of Monte-Carlo search,
downhill-Simplex, and the variable metric method (James and Roos
1977).  Due to the optical problems inherent in pre-refurbishment HST
data and the high central concentration of light in the galaxies
themselves, we applied the following additional adjustments to the
minimization procedure.

Even with an accurate point-spread function (PSF) and careful image
deconvolution, the effect of the spherical aberration of the HST
primary cannot be completely corrected within 0.$^{\prime\prime}$1 of
the galaxy centers (Lauer {\it et al.} 1992b).  Therefore the fitting
domain is restricted to radii greater than 0.$^{\prime\prime}$1.
Nevertheless profiles do differ greatly within 0.$^{\prime\prime}$1,
so there is some information there.  To utilize this, the {\it total
light} within 0.$^{\prime\prime}$1 (as opposed to its distribution)
was used as an additional constraint on the the fit.  The weight of
the central light is equivalent to the sum of weights for the
resampled data points which lie within 0.$^{\prime\prime}$1.  Note
that the constraint given by the central light does not necessarily
force the fit to pass through the data points within
0.$^{\prime\prime}$1.  Large departures between the fit and the
innermost data point often occur.  This is because the total light is
easily adjusted by only a slight change in the surface brightness in
the outer region (but still within 0.$^{\prime\prime}$1), where the
areal coverage is comparatively large.

Nuker law fits to the sample galaxies are shown in Figure 3, and the
structural parameters are listed in Table 1.  For a few galaxies with
nuclei such as NGC 1331, NGC 3599, and NGC 4239, a separate fit was
made to the bulge component only, excluding the nucleus. In these
cases, the central light constraint is also ignored in the fit.  Any
dust patches were masked out from the raw image before the profile
extraction.  Nevertheless the galaxies with significant dust features,
such as NGC 524, show distinct distortions in their profiles.  In
these cases, we set the weights to zero in the regions obviously
affected by dust obscuration. For each galaxy, the fitting domain is
shown by the solid symbols. The RMS deviation computed over the
fitting domain between the model and the original surface brightness
measurements is listed in Table 1.

While the Nuker fits generally match quite well the observed profiles
out to the outermost HST data points ($r \sim 10^{\prime\prime}$),
there are cases like Abell 1020, NGC 4478, and NGC 4486B, where the
outer parts deviate (both upward and downward) from the Nuker law.
Even galaxies which show good agreement with the Nuker law within
$10^{\prime\prime}$ in general will also fail at much larger radii
beyond the field covered by the present HST data, as the profiles
follow a curving de Vaucouleurs law, not a power law there.  For some
of the sample galaxies, the HST data are combined with ground based
photometry covering much larger area and presented in Kormendy {\it et
al.} (1996, in preparation).

\section{Parameter Uncertainties}

\subsection{Error Estimates}

The deconvolution process can introduce systematic errors into
surface-brightness profiles, particularly if use is made of an
inappropriate PSF. Experiments with different PSFs were carried out in
Paper I and revealed both photometric offsets and changes in slope out
to $\sim 2^{\prime\prime}$.  These effects are more pronounced for
galaxies with steep surface brightness profiles near the center.
However, the experiments showed that even for galaxies with
particularly sharp nuclei, the systematic errors in the inner parts
(where the effects of deconvolution are greatest) are small ($\sim
0.^{\rm m}06$). We note that the slope change would be echoed primarily by
$\gamma$, $\alpha$, and $I_b$; $\beta$ and $r_b$ would be essentially
unaffected.

The deconvolution process also amplifies photometric uncertainties,
and we have carried out simulations to better understand the errors in
the Nuker parameters introduced by deconvolution of data with a finite
signal-to-noise ratio. Each simulation was carried out as follows:
({\it i}) Poisson noise appropriate to the number of collected photons
in each pixel was added to the raw (undeconvolved) image of a galaxy.
({\it ii}) Each such realization was deconvolved using 80 iterations
of the Lucy-Richardson deconvolution algorithm.  ({\it iii}) The
two-dimensional data were fitted with a series of concentric ellipses
as described in Paper I.  ({\it iv}) The mean profiles were fitted
with the Nuker law and the resulting parameters stored.

It is important to note that we have added artificial noise on top of
the natural noise already present in the data. In this sense, our
simulations {\it overestimate} (by $\approx\sqrt{2}$) the errors due
purely to S/N. To isolate the effects of S/N alone, we generated a
noiseless model image and repeated the experiments. By comparing the
results of these experiments to those using real data, we were able to
estimate to what extent the goodness of fit was determined by
non-photon statistical errors, such as changes in ellipticity,
flat-field defects, unsubtracted stars, undetected dust obscuration,
{\it etc.}

We chose three galaxies for this exercise : Abell 1831, NGC 4889, and
NGC 4434.  Abell 1831 and NGC 4889 are typical examples of well
resolved core galaxies with small $\gamma$, while NGC 4434 is a
power-law galaxy.  These galaxies also span a large range in central
signal-to-noise ratio. Abell 1831 has the lowest central
signal-to-noise ratio in the sample, with a peak data count around
$1.0 \times 10^3$DN (digital numbers) per pixel. By comparison, NGC
4889 has a peak count of $6.9 \times 10^3$DN per pixel, and NGC 4434
has a peak count of $1.1 \times 10^4$DN per pixel.

For each galaxy one hundred realizations of the original image were
generated.  These realizations were then deconvolved and independently
analyzed.  We constructed a symmetric, two-dimensional, noiseless
model image based on the best-fit parameters for Abell 1831, convolved
with an appropriate PSF, and scaled to match the signal-to-noise ratio
in the real image. Experiments with this model image provided 
a worst-case upper limit on the degree to which the results in
Table 1 are affected by purely statistical uncertainties.

The profiles fitted to the deconvolved, simulated images are shown in
Figure 4.  As expected, Abell 1831 and its model representation show
the largest spread among the three galaxies, and deviations from the
best-fit profile are apparent out to $r \sim 1^{\prime\prime}$.  NGC
4889 also shows an obvious dispersion, but in this case the deviations
are generally confined to $r < 0.^{\prime\prime}1$, and the fit
parameters are rarely significantly different from their best-fit values.
NGC 4434 has the highest signal-to-noise ratio and, despite the
steepness of the profile near the center,  the effects of
photon noise and deconvolution appear to be minimal.

We note that deconvolving a natural image with added noise changes the
shape of the simulated profiles somewhat, at least in the case of
Abell 1831 and its model represenation. There is a slight increase in
surface brightness in the region $ 0.^{\prime\prime}1 < r <
1^{\prime\prime}$.  The Nuker parameters derived from the simulated
profiles therefore are somewhat different from those listed in Table
1.  This effect is not detected in the other two galaxies with higher
signal-to-noise ratios, and we attribute it to the non-conservative
nature of the Lucy-Richardson deconvolution algorithm employed.

The frequency distributions of the best-fit Nuker parameters for the
four galaxy images are plotted in Figure 5.  The mean parameter value
and associated standard deviation in each panel were determined by
fitting a Gaussian to each frequency histogram.  As expected from the
amplitude of the variations apparent in the simulated profiles, Abell
1831 shows the largest standard deviation for each of its Nuker
parameters except $\beta$.  $\beta$, the slope of the outer power-law,
is rather stable in Abell 1831 because $r_b$ is comparatively small
and the outer power law portion of the profile is sampled over a
large radial range.

The average values of the Nuker parameters for Abell 1831 and its
model representation are also not identical. This is due in part to
the different noise characteristics of the two images near the break
radius and their amplification in the deconvolution process. It is
also due to the fact that the model image does not exactly mimic the
real data. For example, the slight dip apparent at $r \sim
0.^{\prime\prime}15$ in the real data is not reproduced in the model
image, and the best fits will differ accordingly.

We note that there is no significant difference in $\chi^2$
between the best-fit Nuker parameters for Abell 1831 and those for its
noiseless model representation.  This implies that our computed
$\chi^2$ is much less dependent on signal-to-noise ratio than it is on
other factors. These factors would include image defects as well as
real discrepancies in form between Nuker profiles and those of
galaxies.

In spite of its low signal-to-noise ratio, both Abell 1831 and its
model representation show that errors in the Nuker parameters
introduced by photon noise and deconvolution are small. We conclude
that the entries in Table 1 are largely unaffected by differences in
signal-to-noise ratio.

\subsection{Covariances}

The uncertainties in the derived parameters and the coupling of the
parameters with one another are usually investigated using the
distribution of $\chi^2$. For any two parameters, we can project the
5-dimensional $\chi^2$ distribution onto the appropriate axes. This
reveals not only the how individual parameters are constrained by the
data, but also the degree to which various parameters are correlated.
However, since the data have been resampled and assigned uniform
weights in order to put equal emphasis on different portions of the
surface brightness profiles, the values of $\chi^2$ we compute, while
relatively correct, are not properly normalized to compute absolute
uncertainties. Our use of the total light within $0.^{\prime\prime}1$
as an arbitrarily weighted, additional constraint exacerbates
this problem. We could conceivably compute an {\it a priori} $\chi^2$
based on photon statistics, but that would not include errors due to
flat-fielding defects, improper deconvolution, and systematic mismatch
betweeen the Nuker law and the galaxies.

In Figure 6 we show projected, constant-$\chi^2$ contours for several
combinations of Nuker parameters for the three galaxies considered in
the previous section.  For the reasons given above, the confidence
limits for each parameter cannot be directly estimated from the
initially computed $\chi^2$ values. To remedy this, we have normalized
the computed $\chi^2$ values using the observed distributions of
best-fit parameters found in the Monte Carlo
simulations. The inner, thick contours in Figure 6 thus correspond to
the 1 $\sigma$ errors determined from the Monte Carlo simulations.
In order to show the gradient of the projected $\chi^2$ distribution,
the outer, thick contours are used to indicate the 10 $\sigma$
level. This Monte Carlo normalization includes photon statistics but
not the other error sources note above, so the treatment is
approximate. 

	The general conclusion from these plots is that, while certain
error pairs are correlated, the only truly serious correlation is
between $r_b$ and $\mu_b$ for power law galaxies. This is to be
expected, as these parameters are formally completely unconstrained
for pure power laws with zero curvature.

If we exclude those galaxies with nuclei (NGC 1331, 3599, 4239, and
6166) or with obvious and widespread dust (NGC 524), the remaining 52
galaxies give a mean RMS deviation between the best-fit Nuker law and
the surface brightness measurements of 2.7\% (0.03 mag). This is
comparable to, though somewhat larger than, an RMS of 1\% obtained by
Gebhardt \etal (1996) using smoothed spline fits to the data. This
simply reflects the fact that the Nuker law is a smoother estimate of
the surface brightness compared with the non-parametric fit.  For the
purposes of determining the physical properties of an individual
galaxy, it is therefore preferable to invert a non-parametric form of
the surface brightness profile. However, if one is searching for
global trends and systematics, the Nuker parameters provide a
convenient basis for comparison. The conclusions reached in Paper I
and Faber \etal (1996) based on the Nuker parameters are not altered
with the application of non-parametric methods.

\section{Summary}

In this paper, we derive a set of structural parameters for 57
early-type galaxies observed with HST.  These parameters are based on
the empirical ``Nuker'' law, which successfully describes the central
surface-brightness distributions within the radius range from
$\sim$0.$^{\prime\prime}$1 to $\sim$10$^{\prime\prime}$ for most
galaxies in the sample. These parameters have been used in Paper I and
Faber \etal (1996) to identify distinct classes of objects,
characterized by significant differences in one or more of the Nuker
parameters.  While these parameters can in principle also be used to
evaluate luminosity densities and (with the availability of suitable
spectral information) the run of velocity dispersion with radius, we
believe this would be better left to non-parametric methods (Gebhardt
\etal 1996), which are more faithful to small but significant
undulations in the observed profiles.

\acknowledgments

This research was supported in part by NASA grant GO-02600.01-87A.

\clearpage

\figcaption{
Surface brightness profiles for $\eta$-models with a range of
$\eta$ values ({\it open circles}).  Solid line represents Nuker law
fits to each model profile.}

\figcaption{
Nuker parameters for $\eta$ models.  Left diagrams are for
the fits made to the radius range from $10^{-5}$ to $10^{3}$.  Right
diagrams are with the restricted radius range of $0.1 < r < 10$,
which is comparable to our fitting regime for the sample galaxies.
$\mu_b$ is the surface brightness at $r_b$, {\it i.e.} $\mu_b = -2.5
\log_{10}I_b$.}

\figcaption{ Mean surface brightness profiles ({\it filled and open
circles}) and Nuker law fits ({\it solid lines}) for the sample
galaxies.  The surface brightness is in $mag/sec^2$. The filled
circles indicate the domain over which the fits are were computed,
while the open circles denote the points excluded from the fit. The
data points inside $r\sim0.^{\prime\prime}1$ are regarded as
uncertain, therefore are not directly used in the fitting (see text).}

\figcaption{
Simulated surface brightness profiles ({\it solid lines}) for Abell
1831, NGC 4889 and NGC 4434, shown with the original profile ({\it
open circles}).  The top diagram is for the noiseless model image of
Abell 1831.  The peak data counts are $1.0 \times 10^3$DN for Abell
1831, $6.9 \times 10^3$DN for NGC 4889, and $1.1 \times 10^4$DN for
NGC 4434 respectively. The means and RMSs of the numerical experiments
at $0.^{\prime\prime}03$ and $0.^{\prime\prime}05$ are shown as open
squares with error bars.}

\figcaption{
Frequency histogram of Nuker parameters derived from 100
simulated profiles of the noiseless model image of Abell 1831, Abell
1831, NGC 4889, and NGC 4434.  A Gaussian distribution is evaluated for
each parameter and shown by solid curve.  The mean and standard
deviation values are given.  $\mu_b$ is in $mag/sec^2$ and $r_b$ is in
arcsec.}

\figcaption{
Parameter cross-correlations shown by the $\chi^2$ ellipses
for Abell 1831, NGC 4434, and NGC 4889.  $\mu_b$ is in $mag/sec^2$ and
$r_b$ is in arcsec.  The thick solid curves indicate 1$\sigma$ and
10$\sigma$ ellipses as scaled by the results of Monte-Carlo simulations
given in Figure 5.  The locations of the minimum $\chi^2$ are indicated
by crosses.}

%\clearpage

%{\sc TABLE} 1. Structural Parameters.

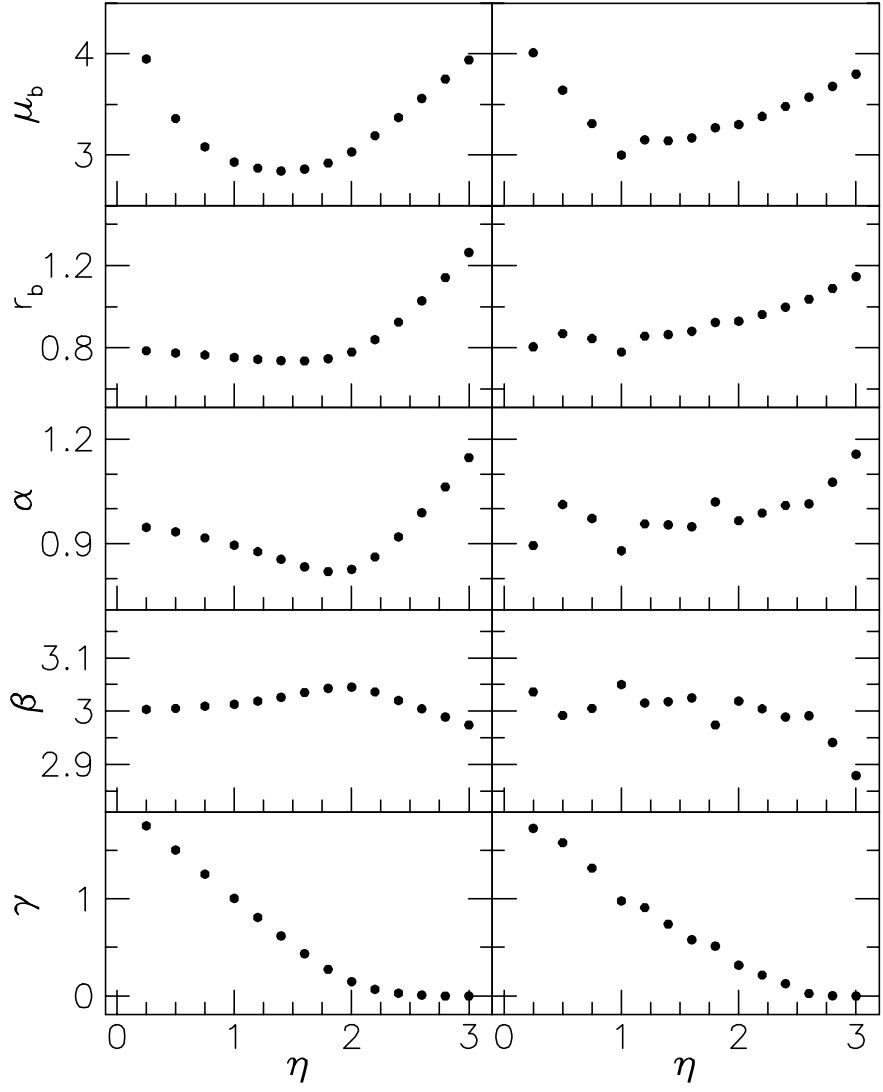
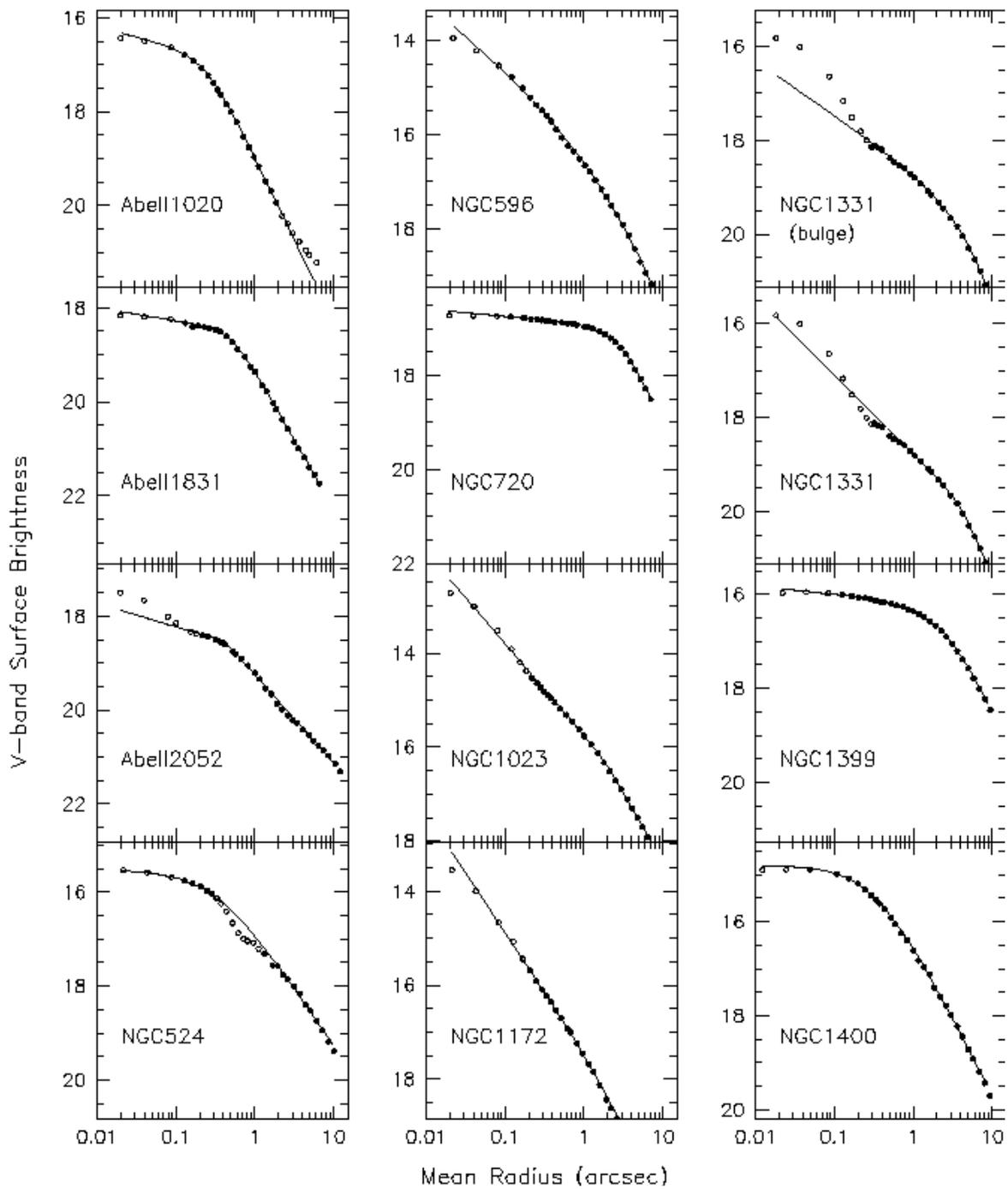
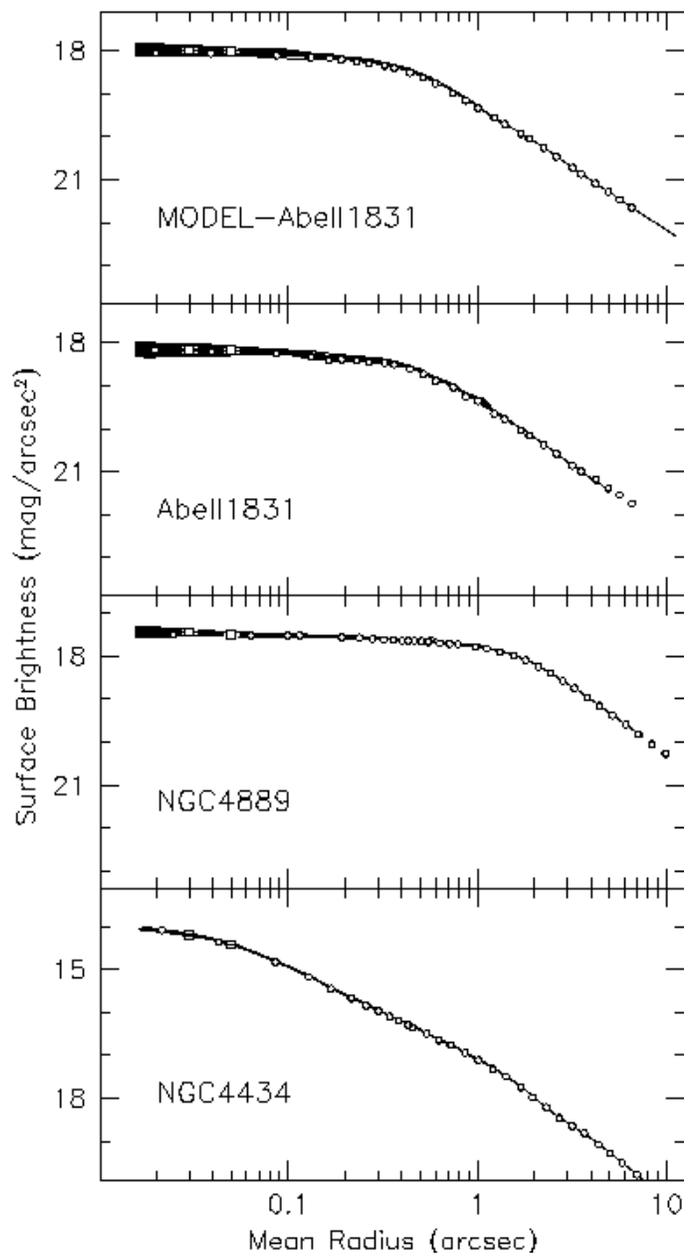
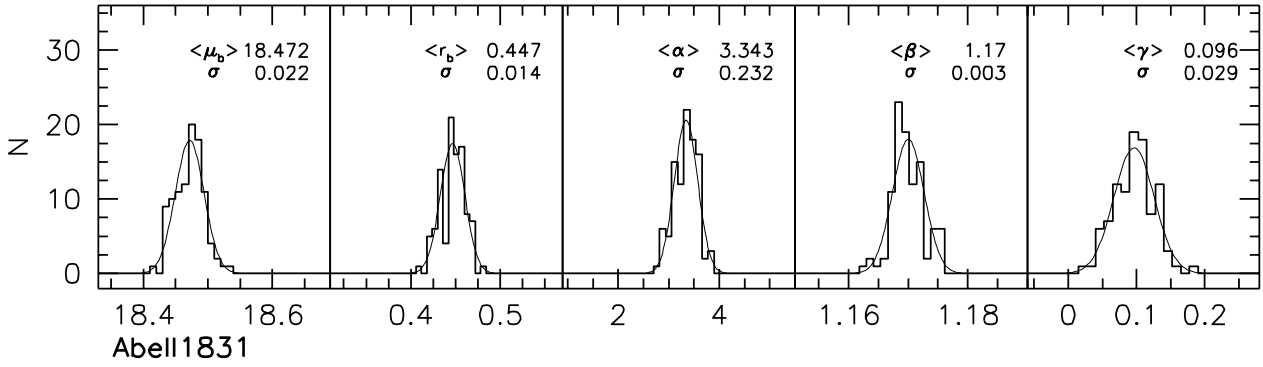
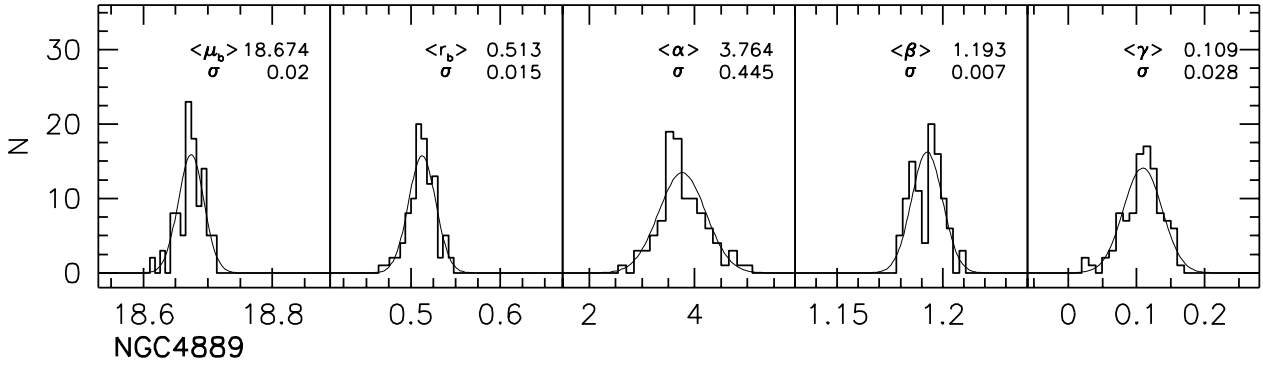
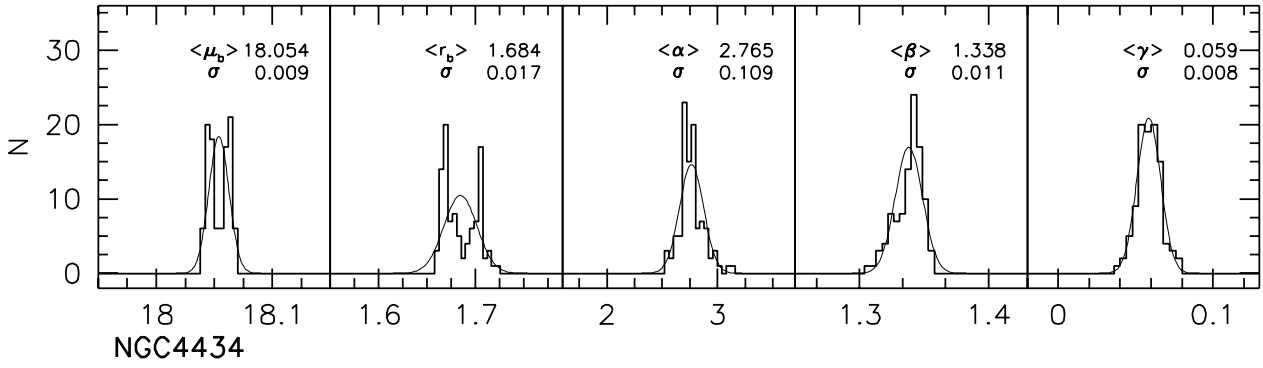
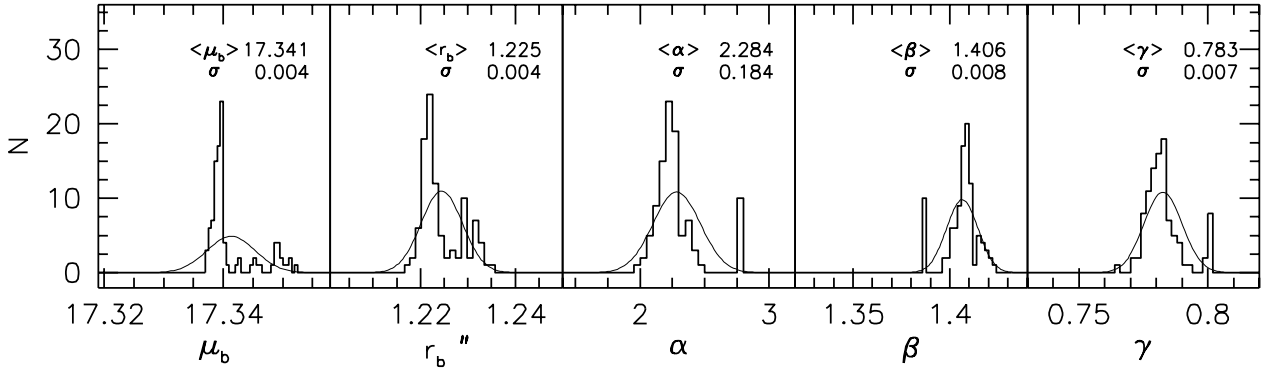
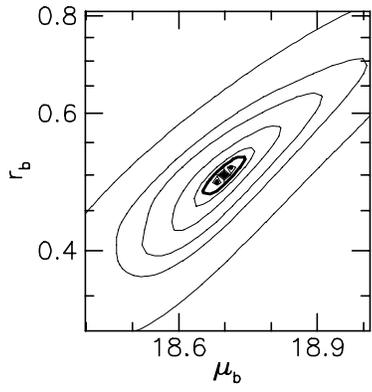
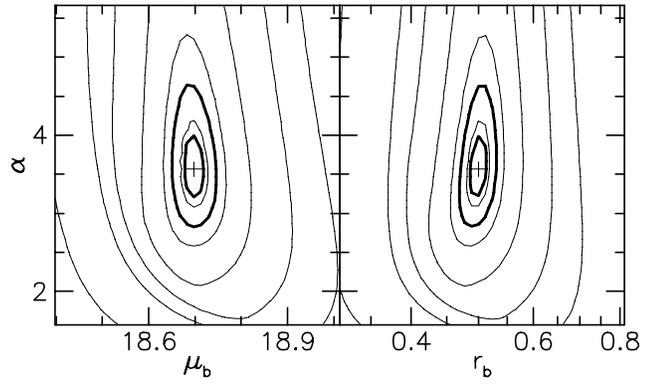
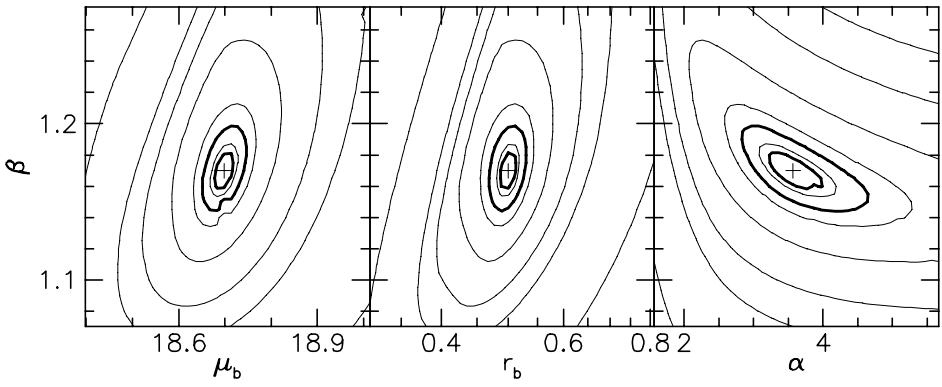
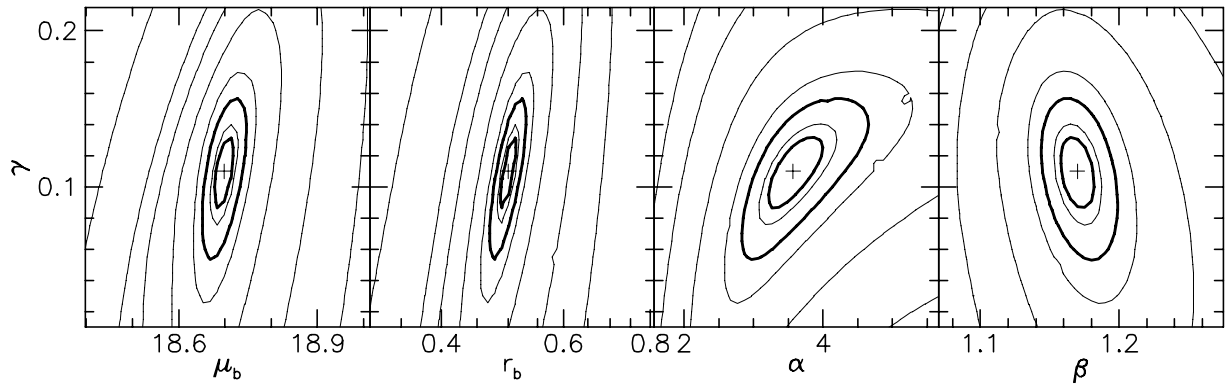
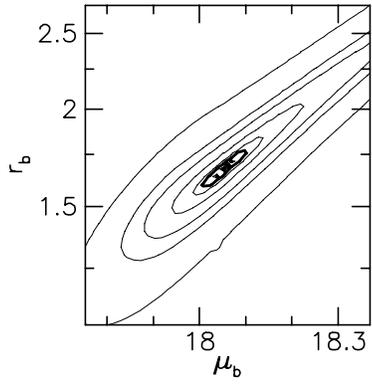
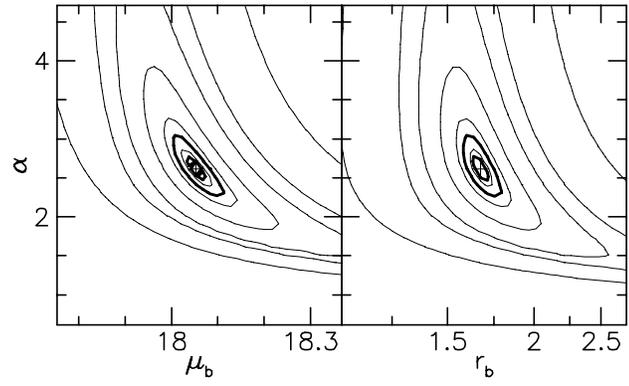
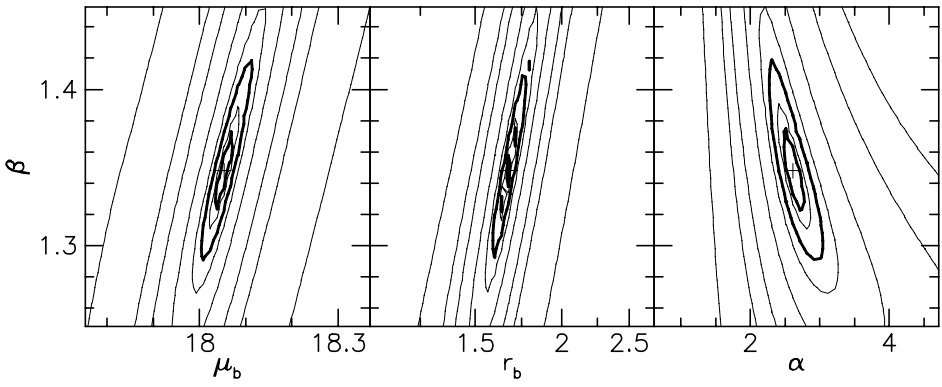
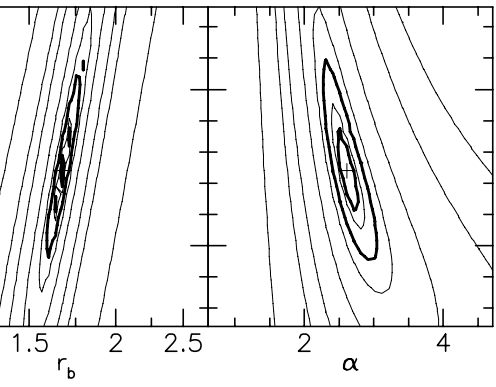
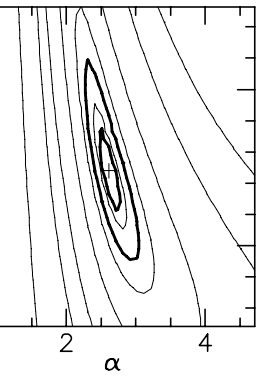
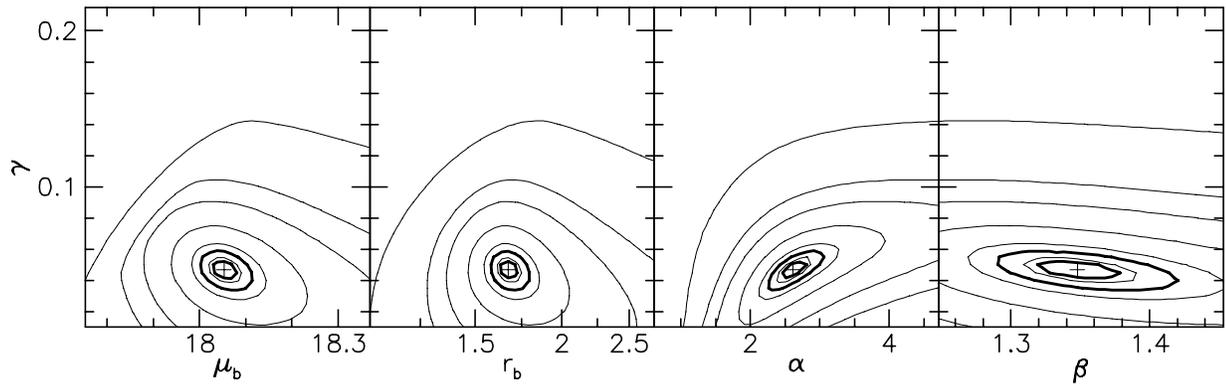
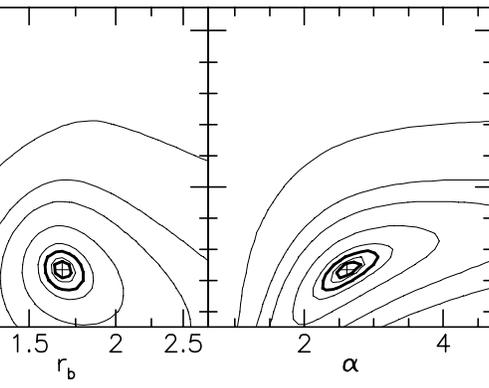
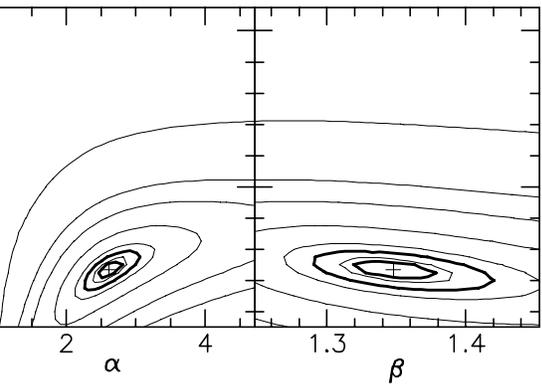
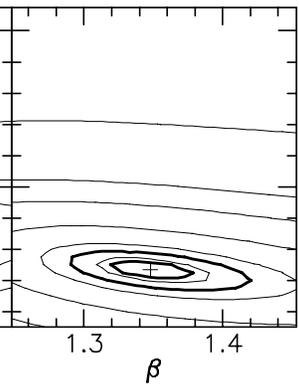
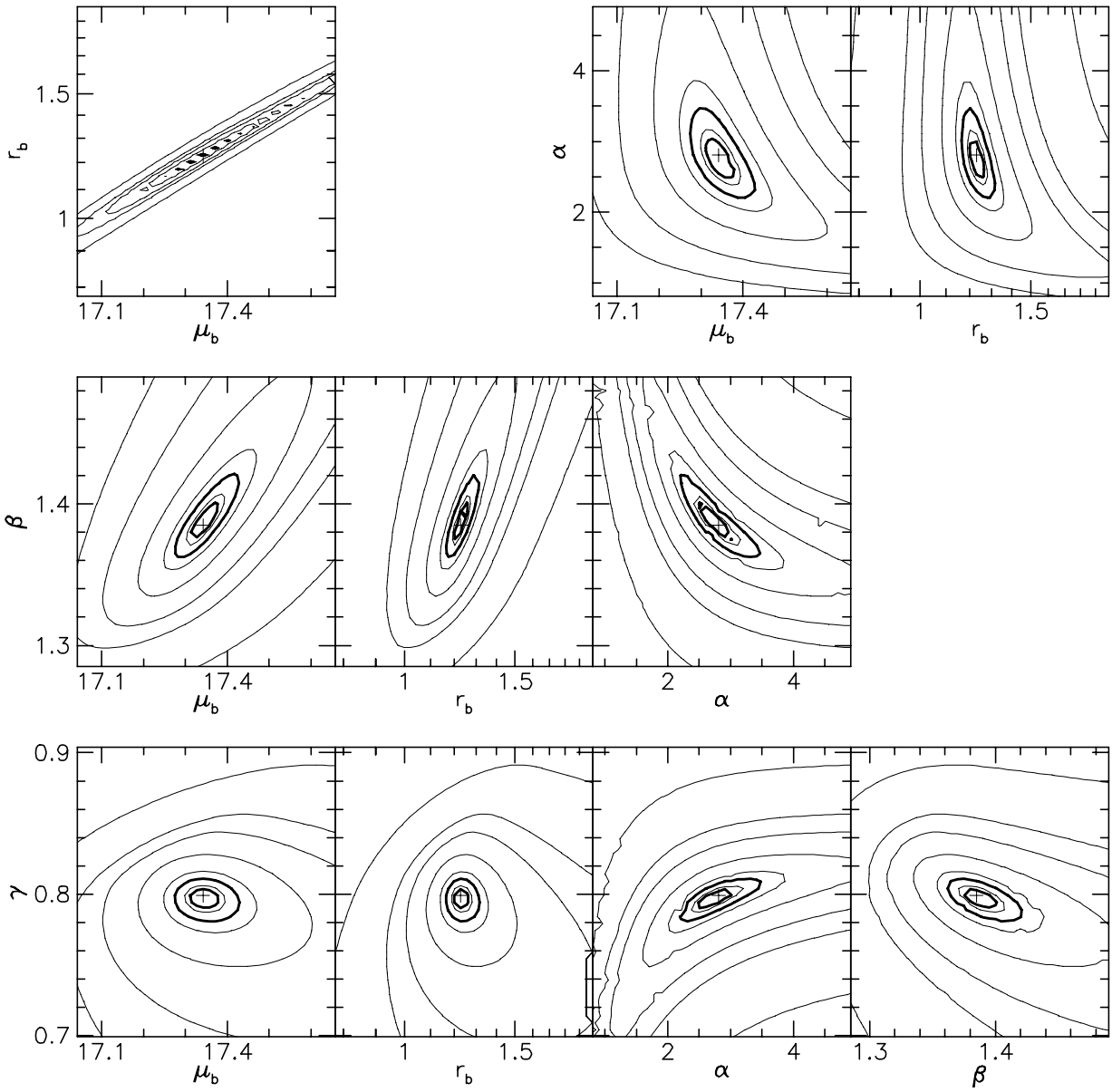
\end{document}

\egtrm

\centerline{TABLE 1 : STRUCTURAL PARAMETERS}

\renewcommand{\arraystretch}{0.8}
\begin{tabular}{lrrrrrrrl}
\noalign{\vspace{5mm}}
\hline \hline \noalign{\medskip}
Galaxy&Sample
&\multicolumn{1}{c}{$\mu_b^\dagger$}&\multicolumn{1}{c}{$r_b$}&\multicolumn{1}{c}{$\alpha$}&\multicolumn{1}{c}{$\beta$}&\multicolumn{1}{c}{$\gamma$}&\multicolumn{1}{c}{RMS (\%)}&\multicolumn{1}{c}{Note~~~~~~~~~~~~~~~~~~~~}\\
\noalign{\smallskip}
\hline
\noalign{\medskip}
Abell 1020 & 3912 & 17.160 & 0.\hspace{-0.8mm}$^{^{\prime\prime}}$\hspace{-1.1mm}236 & 2.564 & 1.394 & 0.175 & 1.34 &             \\
Abell 1831$\ddagger$ & 2600 & 18.698 & 0.500 & 3.567 & 1.170 & 0.110 & 2.19 &             \\
Abell 2052 & 2600 & 18.600 & 0.419 & 8.022 & 0.749 & 0.201 & 3.71 & nuclear point source excluded in fit  \\
NGC ~~524  & 3912 & 16.121 & 0.322 & 1.293 & 1.001 & 0.005 & 5.36 & dust region excluded in fit             \\
NGC ~~596  & 2600 & 18.116 & 3.501 & 0.758 & 1.968 & 0.554 & 3.37 &             \\
NGC ~~720  & 2600 & 17.500 & 3.208 & 2.315 & 1.660 & 0.062 & 0.51 &             \\
NGC 1023 & 2600 & 16.357 & 1.865 & 4.722 & 1.184 & 0.782 & 2.28 &             \\
NGC 1172 & 3912 & 18.694 & 2.431 & 1.517 & 1.638 & 1.009 & 3.43 &             \\
NGC 1331 & 3912 & 21.035 & 8.157 & 1.378 & 2.842 & 0.472 & 2.22 & nucleus excluded in fit (``bulge'')        \\
NGC 1331 & 3912 & 19.976 & 4.083 & 4.470 & 1.615 & 0.673 & 5.76 & nucleus included in fit        \\
NGC 1399 & 2600 & 17.059 & 3.144 & 1.504 & 1.682 & 0.068 & 0.70 &             \\
NGC 1400 & 3912 & 15.510 & 0.330 & 1.395 & 1.319 & 0 & 3.20 &             \\
NGC 1426 & 3912 & 17.540 & 1.644 & 3.619 & 1.354 & 0.847 & 3.06 &             \\
NGC 1600 & 3639 & 18.951 & 5.133 & 1.255 & 2.177 & 0 & 2.27 &             \\
NGC 1700 & 3912 & 14.041 & 0.088 & 0.903 & 1.303 & 0 & 2.39 &             \\
NGC 2636 & 3912 & 15.708 & 0.093 & 1.838 & 1.140 & 0.042 & 3.11 &             \\
NGC 2832 & 3912 & 17.452 & 0.910 & 1.837 & 1.399 & 0.024 & 1.05 &             \\
NGC 2841 & 3912 & 14.548 & 0.133 & 0.928 & 1.021 & 0.009 & 3.55 &             \\
NGC 3115 & 2600 & 16.248 & 2.906 & 1.466 & 1.429 & 0.791 & 1.59 &             \\
NGC 3377 & 2600 & 12.902 & 0.089 & 1.923 & 1.332 & 0.292 & 2.71 &             \\
NGC 3379 & 3229 & 16.136 & 1.738 & 1.593 & 1.432 & 0.182 & 2.05 &             \\
NGC 3599 & 3912 & 17.390 & 1.219 & 13.760 & 1.670 & 0.580 & 7.43 & nucleus excluded in fit (``bulge'')            \\
NGC 3599 & 3912 & 17.578 & 1.351 & 13.009 & 1.660 & 0.794 & 9.12 & nucleus included in fit             \\
NGC 3605 & 2600 & 17.253 & 0.886 & 9.135 & 1.263 & 0.671 & 2.15 &              \\
NGC 3608 & 3912 & 15.449 & 0.284 & 1.052 & 1.330 & 0 & 1.55 &               \\
NGC 4168 & 2607 & 18.357 & 2.521 & 0.945 & 1.500 & 0.140 & 1.55 &              \\
NGC 4239 & 3912 & 18.028 & 0.815 & 4.237 & 0.967 & 0.352 & 3.20 & nucleus excluded in fit (``bulge'')        \\
NGC 4239 & 3912 & 18.420 & 1.284 & 14.531 & 0.959 & 0.646 & 7.81 &  nucleus included in fit             \\
NGC 4258 & 1038 & 13.712 & 0.046 & 1.047 & 0.873 & 0.002 & 3.74 &              \\
NGC 4365 & 2607 & 15.448 & 1.671 & 2.056 & 1.270 & 0.148 & 1.72 &              \\
NGC 4387 & 2600 & 18.993 & 4.423 & 3.362 & 1.591 & 0.719 & 2.51 &              \\
NGC 4434$\ddagger$ & 2600 & 17.342 & 1.230 & 2.809 & 1.385 & 0.799 & 2.86 &             \\
NGC 4458 & 2600 & 14.541 & 0.118 & 5.265 & 1.427 & 0.490 & 2.88 &              \\
NGC 4464 & 2600 & 17.348 & 1.212 & 1.642 & 1.683 & 0.884 & 1.95 &             \\
NGC 4467 & 2600 & 19.982 & 3.243 & 7.518 & 2.133 & 0.981 & 5.68 &             \\
NGC 4472 & 3229 & 16.659 & 2.398 & 2.031 & 1.165 & 0.039 & 1.31 &              \\
NGC 4473 & 2607 & 17.335 & 3.704 & 0.665 & 2.447 & 0.014 & 1.92 &  double nucleus?            \\
NGC 4478 & 2607 & 15.458 & 0.175 & 3.321 & 0.836 & 0.427 & 2.54 &              \\
NGC 4486 & 1105 & 17.923 & 7.606 & 2.816 & 1.394 & 0.246 & 0.62 &  point source removed; profile from Lauer {\it et al} (1992a)\\
NGC 4486B & 3912 & 14.976 & 0.181 & 2.781 & 1.329 & 0.139 & 1.23 &             \\
%NGC 4550 & 2607 & 16.921 & 0.515 & 33.778 & 0.644 & 1.100 & counter-rotating disk galaxy  \\
NGC 4551 & 2600 & 18.922 & 3.862 & 2.944 & 1.233 & 0.802 & 2.65 &             \\
NGC 4552 & 3286 & 15.512 & 0.649 & 1.483 & 1.297 & 0 & 4.37 &             \\
NGC 4564 & 2607 & 15.722 & 0.516 & 0.255 & 1.905 & 0.047 & 3.35 &             \\
NGC 4570 & 2607 & 17.292 & 2.819 & 3.715 & 1.491 & 0.845 & 2.31 &             \\
NGC 4621 & 3229 & 17.250 & 2.922 & 0.187 & 1.710 & 0.496 & 3.15 &             \\
NGC 4636 & 3912 & 17.732 & 3.213 & 1.644 & 1.329 & 0.129 & 1.36 &             \\
NGC 4649 & 3229 & 17.191 & 3.584 & 2.005 & 1.300 & 0.153 & 2.62 &             \\
NGC 4697 & 3912 & 16.957 & 2.575 & 24.863 & 1.040 & 0.742 & 4.12 &  dust disk excluded from fit            \\
NGC 4742 & 2600 & 16.764 & 1.386 & 48.599 & 1.994 & 1.091 & 4.64 &             \\
NGC 4874 & 3912 & 19.222 & 2.632 & 2.327 & 1.366 & 0.134 & 1.23 &             \\
NGC 4881 & 5233 & 17.330 & 0.498 & 1.763 & 1.358 & 0.757 & 3.25 &    post refurbishment data        \\
NGC 4889$\ddagger$ & 3912 & 18.053 & 1.675 & 2.613 & 1.348 & 0.047 & 1.26 &             \\
NGC 5813 & 3912 & 16.529 & 0.792 & 2.147 & 1.326 & 0.076 & 1.66 &             \\
NGC 5845 & 2600 & 17.625 & 2.273 & 1.266 & 2.742 & 0.515 & 7.61 &             \\
NGC 6166 & 3912 & 19.348 & 2.221 & 3.320 & 0.992 & 0.081 & 1.18 & nuclear point source excluded in fit \\
NGC 7332 & 2600 & 15.803 & 0.767 & 4.250 & 1.343 & 0.897 & 3.24 &             \\
%NGC 7768 & 3286 & 17.092 & 0.397 & 1.916 & 1.210 & 0 &      dust near center        \\
VCC 1199 & 2600 & 19.677 & 1.500 & 7.986 & 1.619 & 1.129 & 3.99 &  VCC galaxies from Binggelli {\it et al} (1985)           \\
VCC 1440 & 2600 & 19.994 & 2.522 & 5.537 & 1.583 & 0.964 & 5.17 &             \\
VCC 1545 & 2600 & 19.710 & 1.205 & 7.652 & 1.018 & 0.625 & 4.69 &             \\
VCC 1627 & 2600 & 20.153 & 2.989 & 2.123 & 2.099 & 0.948 & 2.37 &             \\
\\
\noalign{\vspace{-3mm}} \hline\\
\end{tabular}

$\dagger$ $\mu_b = -2.5 \log_{10} I_b$ ;  mag/arcsec$^2$

$\ddagger$ galaxies used for detailed analysis in this paper